# Physics-informed active learning for accelerating quantum chemical simulations

Yi-Fan Hou, Lina Zhang,† Quanhao Zhang,† Fuchun Ge, and Pavlo O. Dral*

*State Key Laboratory of Physical Chemistry of Solid Surfaces, College of Chemistry and Chemical Engineering, Fujian Provincial Key Laboratory of Theoretical and Computational Chemistry, and Innovation Laboratory for Sciences and Technologies of Energy Materials of Fujian Province (IKKEM), Xiamen University, Xiamen, Fujian 361005, China*

*Email: dral@xmu.edu.cn*

*†Equal contribution*

**Abstract**

Quantum chemical simulations can be greatly accelerated by constructing machine learning potentials, which is often done using active learning (AL). The usefulness of the constructed potentials is often limited by the high effort required and their insufficient robustness in the simulations. Here we introduce the end-to-end AL for constructing robust data-efficient potentials with affordable investment of time and resources and minimum human interference. Our AL protocol is based on the physics-informed sampling of training points, automatic selection of initial data, uncertainty quantification, and convergence monitoring. The versatility of this protocol is shown in our implementation of quasi-classical molecular dynamics for simulating vibrational spectra, conformer search of a key biochemical molecule, and time-resolved mechanism of the Diels–Alder reactions. These investigations took us days instead of weeks of pure quantum chemical calculations on a high-performance computing cluster. The code in MLatom and tutorials are available at https://github.com/dralgroup/mlatom.

## Introduction

The introduction of machine learning potentials (MLPs) pushed the boundaries of what was previously possible in molecular dynamics (MD).[1, 2] MLPs enable simulations of longer time scales and larger systems with higher accuracy.[1, 3-5] Their applications led to uncovering the underlying mechanisms in chemical reactions and accurate *in silico* prediction of physicochemical properties.[2, 4, 6] The construction of MLPs for such applications is, however, not entirely a black box and requires special care when choosing and generating their training data as the quality of MLP is as good or bad as the quality of the data.[7, 8]

In general, the more data available, the more accurate MLP which leads to conflicting requirements for the amount of required data and the computational resources. In an attempt to resolve this conflict, many active learning (AL) procedures were developed to select as few data points as possible for building MLPs of sufficient accuracy[7, 9]. These procedures mostly have the same underlying strategy that involves the selection of the data points via steps of training MLP, running molecular dynamics (MD) with it, selecting points from MD trajectories, labeling these points with the reference quantum mechanical (QM) method, adding them to the training data set and repeating the procedure (Figure 1).[3] The variants of AL depend on the specifics in each step, e.g., instead of MD[10-15], one can use meta-dynamics[16-21] or uncertainty-biased MD[22].

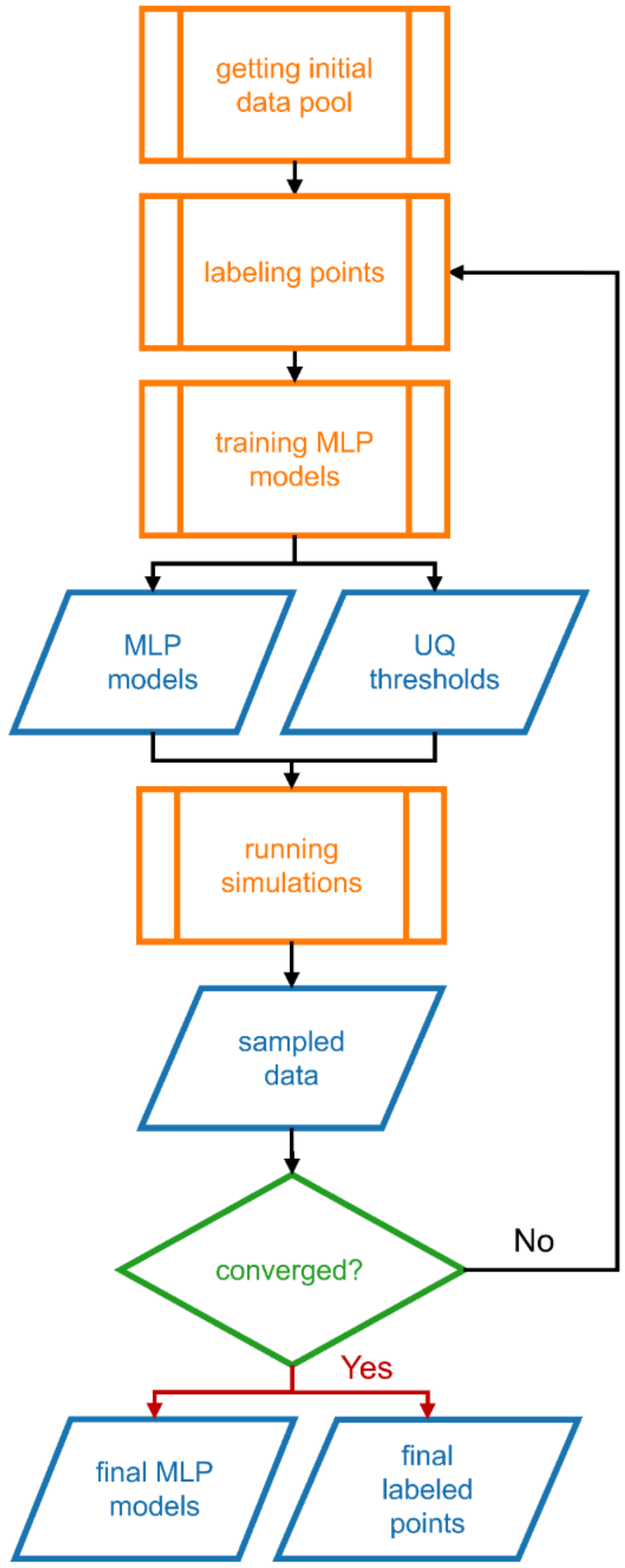

**Figure 1.** Flowchart of active learning for sampling data from potential energy surface and constructing machine learning potentials (MLPs).

The entire AL process hinges on how the points-to-label are selected from MD or its variant. The selection criteria are typically based on statistical or geometric considerations. In the case of statistical consideration, one of the most popular approaches is the query by committee,[23, 24] where several MLPs are trained and the points are chosen when the predictions by MLPs deviate from each other more than a pre-defined threshold. Another statistically-based type of AL uses Bayesian formalism to select points with the highest variance, e.g., derived from MLPs based on the Gaussian process regression[11, 14, 15] or Bayesian neural networks (NNs)[25]. In the case of the geometric criteria, points can be sampled, e.g., when a structural parameter goes beyond the minimum or maximum value in the previous training data set[12, 13].

The approaches based on statistical criteria usually require lots of manual experimentation with the thresholds and sometimes strategies are employed when the AL is started with the large thresholds which are adaptively decreased.[6, 26] The approaches based on geometric criteria require the choice of structural representations[12, 13].

One of the major problems with statistically-based criteria, e.g., based on the query by committee, is that there is no guarantee that the deviation between MLPs would be large from under-sampled regions because MLPs might provide similarly wrong predictions. There is also no single recommendation on how to generate the diverse models in the ensemble and, e.g., models can be trained using different initial weights or on different data splits. Geometric criteria, on the other hand, may miss key conformations that are between the minimum and maximum values.

Here we introduce an end-to-end AL protocol for selecting points based on physical considerations, with automatic selection of the initial data pool and uncertainty quantification (UQ) criteria. We show that this protocol enables building data-efficient, robust MLPs for different applications, from vibrational spectra simulations to the exploration of the conformer space to analysis of the reaction mechanisms. Our implementation starts from the initial molecular structure and ends with the final simulation result by performing all required steps for sampling, labeling, and machine learning in a seamless workflow implemented in MLatom[27, 28].

In these applications, we sample the geometries by performing quasi-classical MD (see Methods). This dynamics represents an additional challenge[29] to MLPs as it account for zero-point vibrational energy and the distribution of the kinetic energies corresponding to relatively high instantaneous temperatures of around thousand Kelvin, i.e., the MLPs have to learn highly distorted structures with a broader potential energy range than in the case of classical MD. The MDs are propagated with the ANI neural network potential which was demonstrated to display a good cost/accuracy balance[30] and was widely employed in related tasks (see Methods)[22, 31, 32].

## Results and discussion

We first overview the key components of the end-to-end AL protocol with specific details given in *Methods*. Then we show several applications of this protocol.

### *Physics-informed sampling*

Our goal is to sample relevant points from the potential energy surface (PES) based on rigorous physical and statistical considerations. The key hypothesis is that, in theory, it is possible to build an MLP ideally representing the PES by sampling points fully recovering PES shape in the region relevant for simulations. This hypothesis is justified by a universal approximation theorem for NNs used in this study. In theory, PES shape can be fully recovered by brute force dense sampling from the grid but, in practice, for realistic multi-dimensional systems, it is impossible due to the curse of dimensionality. Luckily, as research in AL and MLPs manifests, for many real chemical systems it is enough to sample an affordable, relatively small number of points on PES. Thus, the biggest question is how to find these affordable number of points.

According to our hypothesis, the *PES shape must be contained in the sampled points*. If it is not, then the model extrapolates. Consider the 1D-PES example, Figure 2, if we only have two initial points at iteration 0 of AL, then we cannot tell how the function will look between them. The additional physical information provided by energy gradients can help us to extrapolate better and get a pretty good agreement with the true PES in the vicinity of the initial points, but, eventually, the extrapolation ability will deteriorate the farther we go from the initial points. We claim that the only way to have a confident prediction of the PES is to sample additional points when the extrapolation is not working anymore. The question is how to detect the breakdown of extrapolation.

Our innovation is that, in contrast to all previous AL procedures, we detect when the MLP breaks down based on the amount of physical information provided in the already sampled points. We implement a practical realization of the AL protocol based on these considerations by training two MLPs using different amounts of physical information: the main one used for simulations (MD) trained on energies and gradients while the auxiliary MLP (not used in MD) we train only on energies. Both models are trained on the same geometries (i.e., same training/validation splits). The working principle of this implementation is illustrated in a 1D-PES example: our protocol ensures by construction that the main and auxiliary MLPs start to deviate when going away from two initial sampled points and, when the deviation is too large, we sample the new point, label it with the QM method, and retrain the models to obtain refined PES representations in the next iteration 1 (Figure 2).

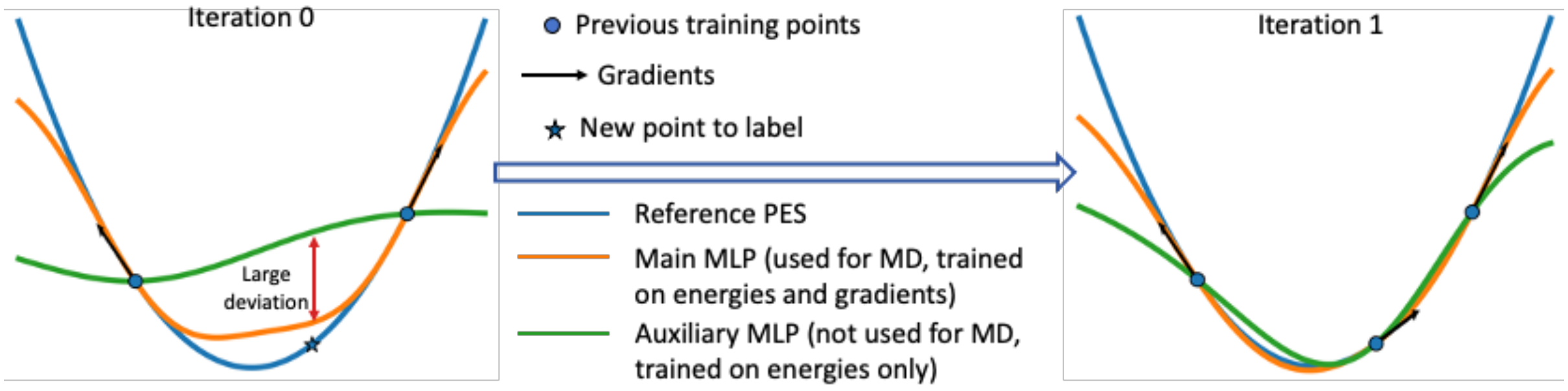

**Figure 2.** Physics-informed potential energy surface sampling illustrated on learning 1D-PES. Blue corresponds to the reference PES, orange – the main MLP trained on both energies and gradients, and green – the auxiliary MLP trained on energies only. Blue circles indicate training points and black arrows show the gradients at training points. In the left panel, the blue star point is sampled where two MLPs give very different predictions, which improves the MLPs as shown in the right panel.

In our scheme, only the main model is used for simulations while the auxiliary model is merely employed to judge whether the main model strays away from the known region, i.e., to detect the breakdown of the extrapolation regime. We do not average the model predictions like in the typical query-by-committee method and only use the main model, which is more accurate due to the inclusion of both energies and gradients, for the simulations. This scheme has a further advantage in that the MD propagation with a single main model is much faster than in the case of AL based on the query-by-committee approach which needs to evaluate several MLPs and take their average at each time step. Similarly, the training time of the auxiliary MLP model is practically negligible compared to the time needed to train the main model.

The practical implementation of the scheme still requires us to make several decisions on how to sample initial data to initialize the AL iterations and what deviation between the main and auxiliary models is too large. We make these decisions based on statistical considerations and automatically select both the initial data and determine the optimal threshold for uncertainty quantification (UQ) defined by the deviation between the main and auxiliary models. This threshold is calculated such that we sample the minimum required points to cover the key regions of the PES to build the near-best-quality MLP. In the following subsections we describe how we achieve it.

***Automatic construction of the initial data set***

To start active learning, we need to generate the initial data set, i.e., choose how we sample configurational space and how many points we include. Since in this work, we use quasi-classical MD, we sample configurational space using the same approach as for generating initial conditions for trajectories, i.e., sampling based on Wigner distribution for a single starting conformer (see Methods). Other initial sampling methods can be used as well.

Once the sampling scheme is chosen, we must determine the number of sampled conformations. For this, we introduce an assumption that the MLP performance for the initial data set will be comparable to the performance on the final data set to be collected by active learning. This allows us to choose the initial training points based on the known statistical behavior of MLP – the error of the potential drops according to the power law with an increasing number of training points (Figure 3a).[33] It means that past some point, increasing the training data will not be worth a small increase in accuracy. Here we choose such a training set that adding $N$ more points would improve the accuracy of the MLP by less than 10%. $N$ is a parameter defined by the user. We gradually increase the number of training points (e.g., add 50 points at a time) and do a 5-fold cross-validation, where the validation error is selected as the measurement of the accuracy of MLP. We fit the learning curve using the equation below:[33]

$$\log(\varepsilon) = \log(a) - b \log N_{\mathrm{tr}} , \tag{1}$$

where $\varepsilon$ is the validation error, $N_{\mathrm{tr}}$ is the number of training points, $a$ and $b$ are coefficients to fit. We stop adding initial points to the training set when the following criterion is met:

$$\frac{\varepsilon(N_{\mathrm{tr}}) - \varepsilon(N_{\mathrm{tr}} + N)}{\varepsilon(N_{\mathrm{tr}})} < 10\%. \tag{2}$$

Otherwise, 50 more points are included in the training set and the procedure repeats. The scheme is shown in Figure 3b. This ensures that in subsequent AL steps, we sample the points with a not-too-bad model.

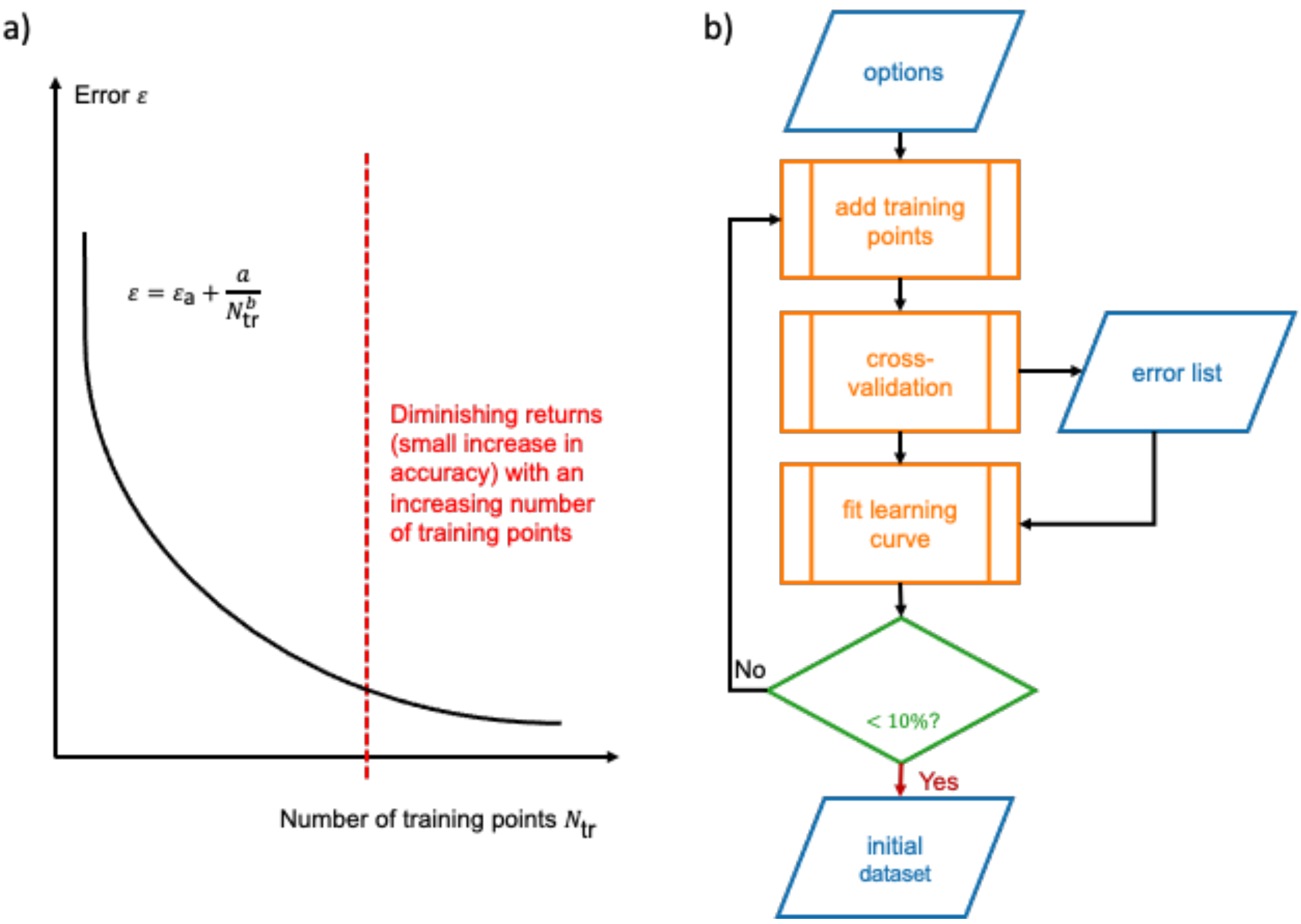

**Figure 3.** Flowchart of initial points sampling scheme (b) based on the known power law for error dependence on the number of training points (a).

### *Automatic determination of the uncertainty quantification thresholds*

A good choice of the uncertainty quantification (UQ) thresholds is also important because we do not want to choose too many bad conformations that would never occur in the actual simulations with the reference method and, at the same time, we do not want to choose too many similar points as it would be wasteful. In our scheme, after we determined the good initial set using the above procedure, we introduce another assumption to choose the UQ threshold based on statistical considerations. We aim to choose such a UQ threshold that at least 99% of the initial data points would be considered confident and assume that this threshold will not be exceeded for the 99% of conformations in the simulation for which the main MLP provides a good description. Thus, we assume that only the points not described well are chosen. In the end, this assumption is ensured by the overall AL workflow as only a small fraction of trajectories (5% by default) will be stopped when the UQ threshold is exceeded – meaning that all geometries in all other trajectory time steps are confident.

We quantify the uncertainty $U$ as the absolute deviation between the energies predicted by the main ($E_{\text{main}}$) and auxiliary ($E_{\text{aux}}$) models trained on 90% of the initial data:

$$U = |E_{\mathrm{aux}} - E_{\mathrm{main}}|. \quad (3)$$

We calculate the UQ threshold by evaluating the uncertainties for the validation set (the remaining 10% of the initial data) using the following expression:

$$\mathrm{UQ}_{\mathrm{threshold}} = M + 3 \cdot \mathrm{MAD}, \quad (4)$$

where $M$ is the median and MAD is the median absolute deviation of uncertainties[34]:

$$\mathrm{MAD} = 1.4826 \cdot \mathrm{median}(|U - M|). \quad (5)$$

The factor 1.4826 assumes that $U$ follows normal distribution, and $M + 3 \cdot \mathrm{MAD}$ ensures a confidence level of 99%. Using the median and MAD instead of the mean and the standard deviation is a more robust way of handling data with outliers.

Molecular dynamics are performed using the main model to explore the potential energy surface. If the uncertainty of the molecule exceeds the UQ threshold, the MD is stopped and the geometry in the last step is sampled.

Ideally, the optimal threshold might be the one determined for the best data set representing PES which we do not have – if we had, we would not need to do AL in the first place. Of course, better approximations might exist but using the Occam's razor principle, we start with a reasonable assumption that the initial data set is good enough to determine the UQ threshold. The ultimate quality of the AL will depend on this assumption and, hence, one might want, for some critical applications, manually reduce the UQ threshold for higher accuracy. However, as we know from statistical considerations (see Figure 3b) this might lead to a substantial increase of the sampled points with only modest improvement in accuracy. Alternatively, the threshold can be increased for faster exploration.

***AL convergence***

At some point, we need to stop AL based on some criteria. Here we consider the AL converged if 95% of the trajectories propagated for the required time do not encounter a single case exceeding the UQ threshold. An important consideration is how many trajectories to run and for how long. How long depends on an application and ideally it should be as long as anticipated in the required production simulations (this anticipated result can be updated iteratively). We initially run 100 trajectories but if the previous active learning iteration samples less than 100 points, we have an option to increase the number of trajectories to roughly sample 100 points in the current iteration. This ensures faster convergence of the active learning and reduces the number of expensive MLP trainings.

***AL for vibrational spectra simulations***

We first performed a well-controlled experiment to directly compare the performance of the MLP with the reference QM simulation of the vibrational (power) spectra obtained from long MD trajectories for the ethanol molecule where we can be sure to sufficiently sample the conformational space (see Methods). Active learning converged after AL sampling of only 962 geometries and took ca. two days on relatively modest hardware (single GPU (RTX3080Ti) and 16 CPUs (AMD EPYC 7302@3.0GHz)). The procedure yielded the main MLP trained for ethanol on the reference B3LYP/6-31G* energies and gradients for these geometries.

We used this MLP to simulate the vibrational spectrum by propagating the 200-ps-long dynamics in the NVE ensemble from the initial geometries and nuclear velocities not seen in AL to test the procedure's transferability (Figure 4a). To test how the same model performs for a different distribution of the geometries, we also apply MLP in a rather different regime of room-temperature classical dynamics in the NVT ensemble (Figure 4c). For comparison, we also simulated the reference spectrum from the 100-ps-long dynamics started from the same test initial conditions but using the reference QM method for calculating gradients. The agreement between the MLP and QM spectra is very good, the peak positions are good and the deviation in intensities is well within the uncertainty of the reference QM spectra.

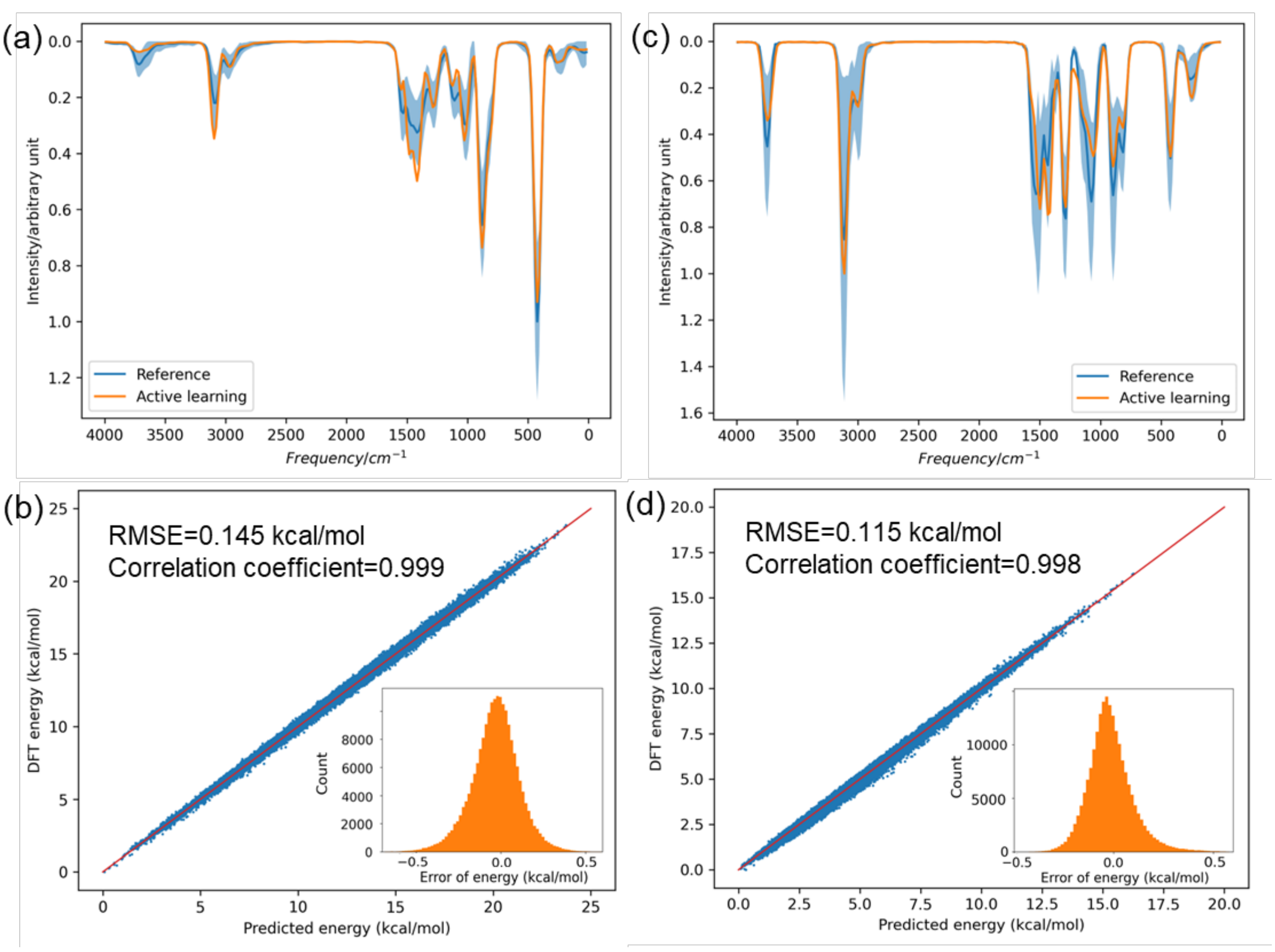

**Figure 4.** Performance tests of the MLP for the ethanol molecule. a, c) Vibrational (power) spectra obtained from the MD in the NVE and NVT ensembles, respectively. MLP spectra are shown in orange and the reference B3LYP/6-31G* in blue. Uncertainties of reference QM trajectories are shown as blue-shaded regions corresponding to one standard deviation of intensities from a hundred trajectories. b, d) Correlation between the MLP-estimated and reference QM energies evaluated on one NVE and NVT reference test trajectories, respectively; the root-mean-squared error (RMSE) and the correlation coefficient are shown together with the histogram of energy errors.

Even for this relatively simple system, the reduction in the required resources is impressive: it took us eight days to propagate the QM trajectory (on 32 CPUs (Intel(R) Xeon(R) Gold 6226R CPU @ 2.90GHz)) and required 200 thousand QM calculations; propagation with MLP took only one hour for a twice-longer trajectory (on RTX3080Ti GPU) after the aforementioned modest investment in resources during the active learning. The accuracy of the MLP in energies evaluated for all snapshots in the test reference trajectory is also quite impressive (Figure 4b). The root-mean-squared error (RMSE) in energy is just 0.1 kcal/mol – much better than reported in previous benchmarks on ANI-type networks for the ethanol PES[30]. Our procedure also yields robust MLP which can be used for long simulations without unphysically breaking down the molecule as is often observed in MD with MLPs.[35-38] These tests show that the physics-informed AL samples important PES regions and ultimately yields data-efficient MLPs with high accuracy and robustness.

***AL for conformational space search***

Our next experiment is designed to test whether the physics-informed AL can also properly sample the complex conformational space. Glycine is very suitable for this because it is well-studied and known to have quite a complex space with eight conformers.[39] We performed active learning based on propagating one hundred 2-ps-long quasi-classical MD trajectories for glycine starting from a single random conformer and then at each active learning iteration, we located distinct conformers in trajectories and restarted dynamics from initial conditions collected from every conformer in the next iteration (see Methods). This procedure was able to identify all eight conformers (Figure 5 and Table S2 in Supplementary Information) after sampling 679 geometries in 6 iterations (6 hours). The AL converged after sampling 1737 geometries (4 days). The conformer search could have taken millions of QM single-point calculations, equivalent to 11 days on one CPU.

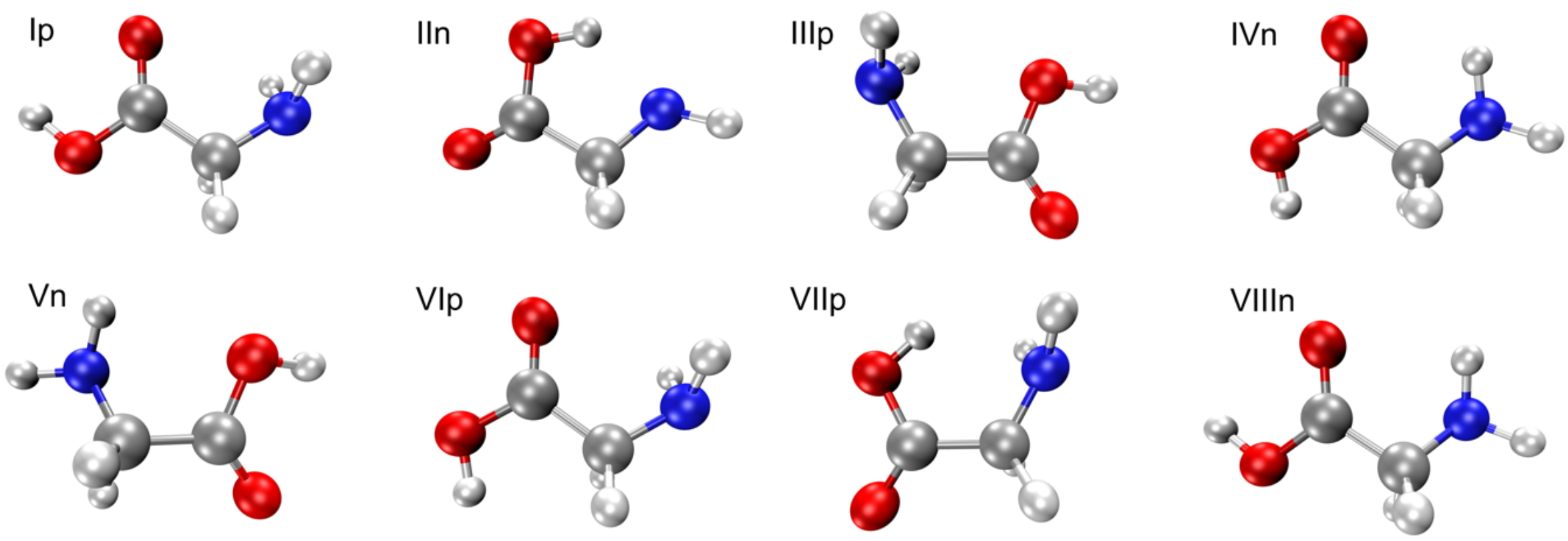

**Figure 5.** Conformers of glycine optimized with the MLP model obtained from AL. The conformer numbering is taken from Ref. [40].

The ML model gives a good representation of the PES as we can see from the internal coordinates scan in Figure 6. Although it is not perfect and deviations might be as large as 0.5 kcal/mol the quality of PES is sufficient for the goal of finding conformers via unbiased quasi-classical trajectories. Some applications such as diffusion Monte Carlo might require higher accuracy which would necessitate choosing tighter thresholds for AL and considering other, more appropriate sampling procedures. Coincidentally, our errors for the PES scans (none of which was part of the training set) are similar to the previous study[22], where the same ANI-type MLP was trained on data obtained through the biased MD for glycine based on the query-by-committee uncertainty quantification. That study did not claim to find all the conformers and the authors mentioned the need for automatic selection of uncertainty criteria – the challenge we answer in this work.

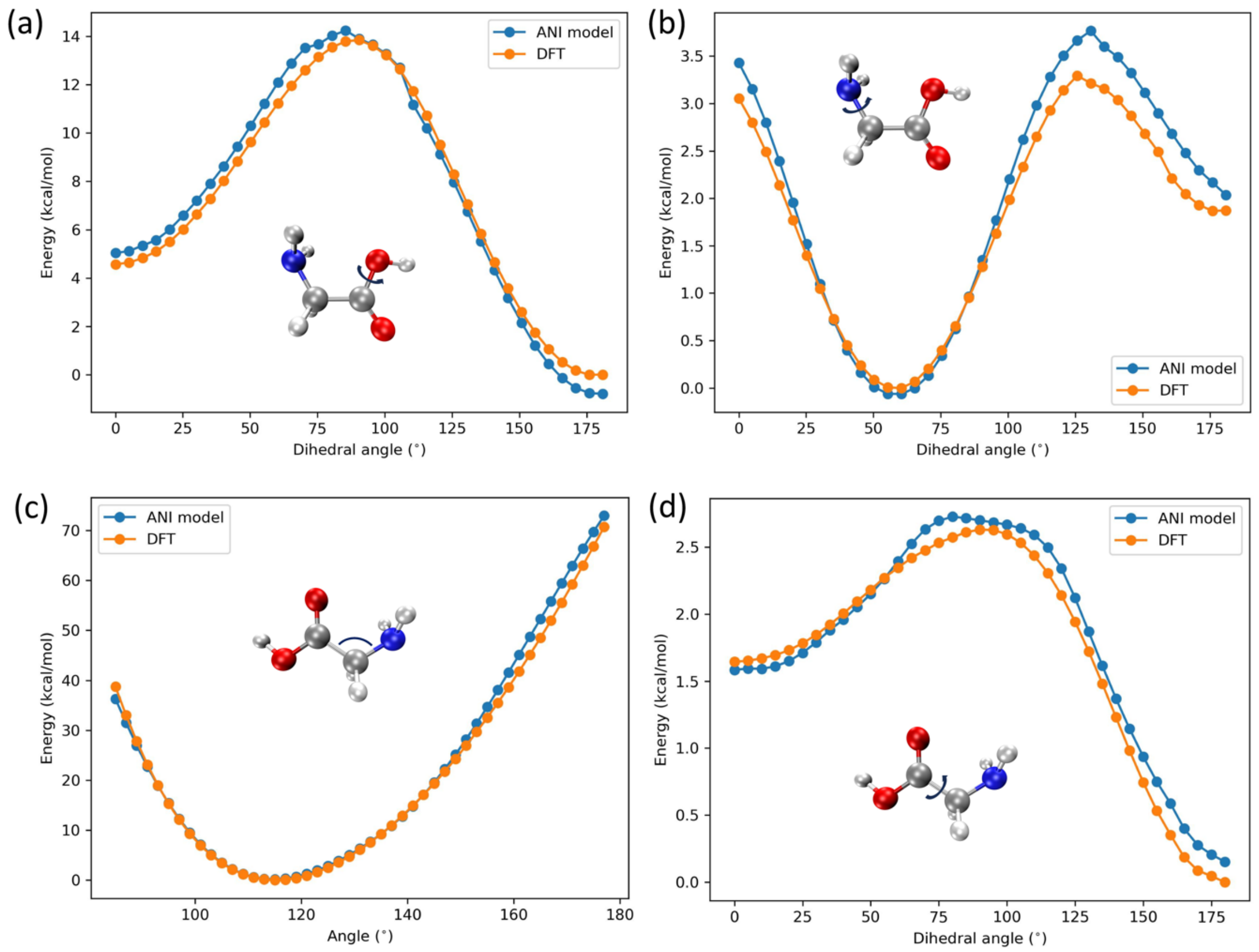

**Figure 6.** Relaxed scan of glycine internal coordinates using B3LYP/6-31G* (orange) and MLP model (blue). Note that we did not make any effort to sample these points and it shows the quality of the MLP model taken from the AL 'as is'.

### *AL for time-resolved reaction mechanism investigation*

Our final validation of the procedure involved the test of how well it performs for the exploration of the PES near the transition state (TS) region. Quasi-classical dynamics started from the initial conditions sampled from the normal modes orthogonal to the direction of the mode with the imaginary frequency, is an informative method for exploration of the intricacies of the dynamic behavior of the reactive events.[41] It is, however, very expensive, as it requires propagating hundreds (when affordable, thousands) trajectories and a recent state-of-the-art study reported the need to extensively use high-performance computing (HPC) clusters as one trajectory required up to a week on 16 CPUs.[42] We aim to drastically slash down these requirements while ensuring the high quality of the simulations.

To calibrate our procedure, we take the textbook Diels–Alder reaction of 1,3-butadiene with ethene which was extensively studied experimentally and theoretically, also with the quasi-classical dynamics.[43] The previously reported TS is symmetric with interatomic distance for forming bonds at 2.272 Å (Figure 7a) and the earlier quasi-classical MD study with QM method UB3LYP/6-31G* also confirmed the concerted nature of the reaction at 298 K.[43] We used the same QM method for the physics-informed AL to be able to compare with the literature.

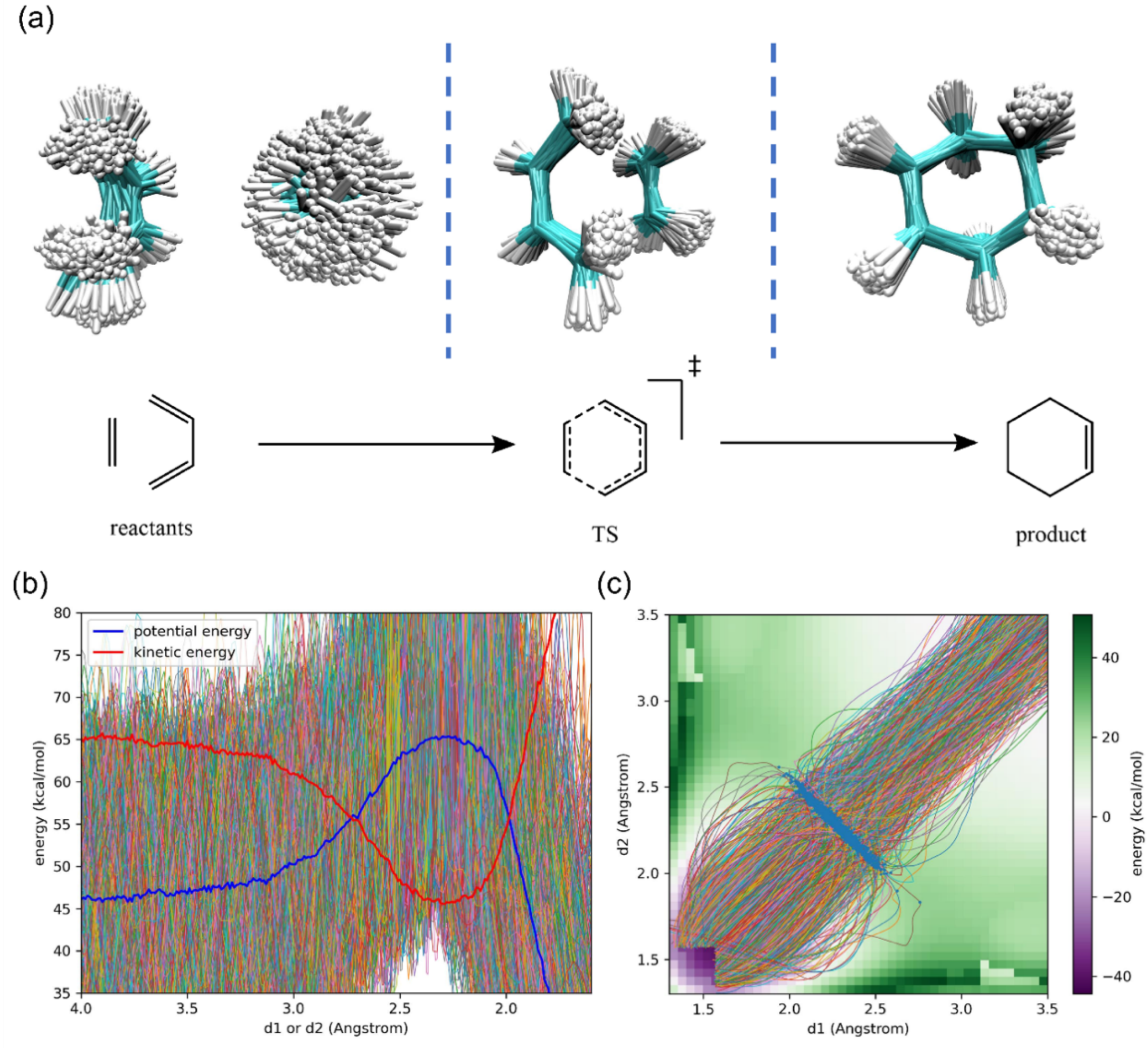

**Figure 7.** Diels–Alder reaction of 1,3-butadiene with ethene. Quasi-classical molecular dynamics was run at 298 K. a) Reaction scheme of Diels-Alder reaction of 1,3-butadiene with ethene calculated using UB3LYP/6-31G* and superposition of reactants, TS, and product geometries from 944 reactive trajectories. b) Potential energy versus forming C–C bond length for 1000 trajectories. The median potential energy (blue) and kinetic energy (red) are also shown. c) Lengths of two forming C–C bonds in Diels–Alder reaction between 1,3-butadiene and ethene. The contour plot shows the relaxed scan of two C–C bonds. Blue points represent initial conditions. The potential energies in b) and c) are relative to the electronic energies of optimized reactants.

Only after 2 days on one single node, our AL procedure produced the MLP model which was used to propagate a thousand 150-fs forward and backward trajectories in under 30 minutes. For comparison, we also ran 128 pure QM quasi-classical trajectories (same as in the previous study[43]) on many computing nodes, where each trajectory took roughly 30 minutes on a single node. We perform the same analysis on these trajectories as reported previously.[43] The key quantitative results are the same between MLP and QM: both predict similar average time to traverse the transition zone with the average time gap between forming the first bond and the second bond of ca. 3.6 fs (Table 1). These results are also very similar to the reported numbers in the independent study[43] but were obtained only at a fraction of the cost with MLP without the need for HPC.

The big advantage of our procedure is that we could afford to substantially increase the precision of the simulations by propagating thousands rather than ca. a hundred of trajectories. While for this study the results are similar at different precision levels, it was recently clearly pointed out that a hundred trajectories are often insufficient for obtaining precise results and in the last five years the reported studies used too few trajectories.[44]

Our more precise results for the Diels–Alder reaction confirmed that, indeed, the reaction is symmetric as both bonds form roughly at the same time in all 944 reactive trajectories (Figure 7c, see Methods for definition of reactive trajectories). We also confirm the same observation as before[43] that this dynamics strongly supports the applicability of the transition state theory for this reaction as the transition state zone represents the major bottleneck as exemplified by the clear potential energy maximum and corresponding kinetic energy minimum (Figure 7b).[43]

**Table 1.** Time to traverse the transition zone and time gap of C–C bond formation for Diels–Alder reaction of 1,3-butadiene and ethene at 298 K. The bond forming time and the dissociating time are also shown. Medians are given in parentheses and standard deviations are also shown.

| | Average time to traverse the transition zone (fs) | Average time gap of C–C bond formation (fs) | Number of trajectories: reactive/total |
|---|---|---|---|
| DFT[43] | 52.6 (50.9) ± 12.0 | 3.9 (3.4) ± 2.9 | 117/128 |
| DFT | 54.0 (52.0) ± 12.9 | 3.6 (3.0) ± 2.9 | 124/128 |
| MLP | 55.0 (52.5) ± 14.3 | 3.9 (3.0) ± 3.7 | 944/1000 |

## Conclusions

We presented the active learning protocol for constructing machine learning potentials based on physics-informed sampling from PES. The key idea is to use different amounts of physical information about the PES shape available in the sampled points. We also proposed an automatic determination of the uncertainty threshold to avoid the manual experimentation of the optimal threshold for the end-to-end AL protocol, which takes the initial molecular structure as input and outputs the final model and simulation results. The AL implementation is based on the open source MLatom's software ecosystem[27, 28] enabling seamless execution of all required steps including labeling with the reference QM method.

We showed the efficiency of this end-to-end physics-informed AL in PES sampling and obtaining accurate MLPs in different application scenarios, *i.e.*, spectra simulations, conformer search, and elucidating the time-resolved mechanism of a Diels–Alder reaction.

Our AL protocol provides a robust solution for speeding up quantum chemical simulations and allows us to break through the bottleneck of expensive MD simulations to make them possible at an affordable cost on the commodity hardware. Indeed, we already used this procedure to discover surprising roaming dynamics phenomena in the Diels–Alder reaction of fullerene $C_{60}$ with 2,3-dimethyl-1,3-butadiene which would otherwise be impossible to uncover with the slow electronic structure calculations.[45] While the presented results were shown for the ground-state dynamics, we are also currently extending the protocol to the surface-hopping excited-state dynamics which are even more expensive.

## Methods

### *Active learning*

Our active learning workflow consists of the following steps (Figure 1):

1. Getting initial data pool.

   In general, the user can provide any initial data pool (collection of geometries). Here we introduce automatic initial pool generation followed by automatic determination of the sampling criteria based on uncertainty quantification as described in the main text.

2. Labeling points by running single-point calculations with the chosen QM method to generate energies and energy gradients.
3. Training MLP models.

4. Running simulations with MLP.

   Implementation is general, but here we run quasi-classical MD.
5. Sampling new points to label from simulation data.
6. Add new points to the data pool.
7. End, if the number of new points is fewer than the pre-defined threshold. Otherwise, repeat starting from step 2. In this work, the threshold is 5% of the number of MD trajectories.

All of these steps are implemented in a seamless workflow with open-source MLatom's Python API[27, 28], which supports all required functionality, i.e., generating initial conditions for MD, labeling points (by running single-point calculations with many supported QM methods), training MLP models of different types (not just ANI and also including their hyperparameter optimizations if required), propagating MD, and handling data (dumping, loading, converting, splitting, etc.). This workflow is integrated into MLatom and will be included in one of its future releases.

***Quasi-classical molecular dynamics***

In this work, we use the quasi-classical molecular dynamics because it accounts for the zero-point energy (ZPE) in contrast to classical dynamics. We follow the reported protocols[41]: generate initial conditions (geometries and velocities) from the normal mode sampling to account for ZPE (details described below) and propagate classical MD trajectories from these initial conditions. In this work, we use velocity Verlet algorithm to integrate the Newton's equations of motion.

***Sampling of initial conditions***

Initial conditions are required for both quasi-classical MD trajectories and for automatic construction of the initial data set (in the latter case no velocities are needed). For local minima, we sample them from the non-sharp Wigner distribution as implemented in Newton-X and described in this software's paper (Eq. 20)[46]. This approach is used without modification for the energy minima (ethanol and glycine simulations).

We adopt another sampling method that was previously developed for transition states.[47] For all the real normal modes that are perpendicular to the reaction coordinate, the quantum number $n_i$ of normal mode $i$ is first sampled from a harmonic quantum Boltzmann distribution[48]:

$$p(n_i) = e^{-n_i h v_i / k_\mathrm{B} T}\left(1 - e^{-h v_i / k_\mathrm{B} T}\right),$$

where $h$ is the Planck constant, $v_i$ is the vibrational frequency, $k_\mathrm{B}$ is the Boltzmann constant and $T$ is the temperature. The coordinates $Q_i$ and momenta $P_i$ are then generated by[49-51]

$$Q_i = A_i \cos(2\pi R_i),$$

$$P_i = -\omega_i A_i \sin(2\pi R_i),$$

where $\omega_i = 2\pi v_i$, $A_i = \sqrt{2E_i}/\omega_i$, $E_i = (n_i + 1/2) h v_i$ and $R_i$ is a uniform random number on $[0, 1]$. The reaction coordinate (corresponding to the normal mode with the imaginary frequency) is fixed while the mass-weighted momentum $P$ is calculated by[52] $\pm\sqrt{-2k_\mathrm{B} T \ln(1 - R)}$, where $R$ is also a uniform random number on $[0, 1]$.

***Ethanol***

Initial training set in active learning and initial conditions of molecular dynamics are generated by Wigner sampling at 300 K. 50 points are sampled in iteration while constructing the initial data for AL and the value of $N$ is set as 50. 100 molecular dynamics trajectories are generated in each AL iteration and run in the NVE ensemble. The propagation time is constrained to be less than 5 ps. A maximum of 50 points are sampled in each iteration, otherwise, the excess points (which are not labeled) are removed randomly. The AL procedure is considered converged if the number of sampled points is less than 5. The UQ threshold is calculated only once and remains unchanged after the very first iteration.

B3LYP/6-31G* is chosen as the reference method. For MD in the NVE ensemble, initial conditions of 100 reference trajectories are generated by Wigner sampling at 300 K. The trajectories using the reference method and the final ML model are propagated with molecular dynamics at 300 K in the NVE ensemble. The first 3 ps of trajectories are removed. The power spectra are calculated by averaging spectra calculated for each trajectory, followed by the normalization of their Riemann sum:[53]

$$\sum_i u_i \mathrm{d}u \overset{!}{=} 1,$$

where $u_i$ is the intensity of the $i$-th point and d$u$ is the distance between the point $i$ and $i$+1.

MD trajectories in the NVT ensemble are propagated in the Nosé–Hoover thermostat at 300 K and using 0.5 fs time step. The first 3 ps of trajectories are removed before calculating the spectra. We only have one 100-ps long reference trajectory and one 200-ps-long MLP trajectory. Thus, for calculating power spectra, each trajectory is split into 10-ps trajectory segments, and the power spectra are calculated by averaging spectra calculated for each segment. The normalization of their Riemann sum is also applied.

***Glycine***

B3LYP/6-31G* is chosen as the reference method. The settings of AL for glycine are basically the same as those for ethanol except that we sample all possible points in each iteration (i.e., up to 100 points). The Wigner sampling is based on all the conformers that are found during active learning. As for the conformer search, we start from one of the conformers. After new ML models are trained (except for the first iteration), the geometries sampled from the last iteration are optimized with the main model. Root-mean-square deviation (RMSD) between optimized molecules and conformers is calculated to measure the similarities; reflections and atom permutations are taken into account. If the smallest RMSD is larger than 0.125 Å, the geometry is considered a potentially new conformer and further optimized by the reference method. If the QM-optimized geometry meets the RMSE criterion, it is considered a true conformer.

***Diels–Alder reaction***

Following the literature[43], UB3LYP/6-31G* is used as the reference method. AL starts from the transition state of the reaction. The initial points and conditions are sampled at 298 K. To increase the sampling efficiency, the number of molecular trajectories is calculated by dividing the maximum number of sampled points (100 points here) by the fraction of sampled points in the previous iteration. This makes the AL sample roughly 100 points in each iteration and the AL is considered converged if the fraction is less than 5%. Following the literature, the propagation time of molecular dynamics is set to 150 fs with time step of 0.5 fs.

Trajectories are propagated in sets of forward and reversed directions, where the initial conditions are the same except that the velocities directions are the opposite. Following the literature[43], the product is considered formed if both forming C–C bonds are shorter than 1.6 Å and two reactants are separated if both bonds are longer than 5.0 Å. If the set of the forward and reversed trajectories leads to the set of reactants and product, they are labeled as reactive; if both trajectories in the set lead to one side of the reaction, they are considered unreactive. Each trajectory discussed in the main text comprises the forward and backward trajectories. 1000 trajectories are generated using the AL model, and 128 trajectories are generated using

DFT. The transition zone covers C–C bond length from 2.02 to 2.52 Å, which includes 98% of the initial points which is the same as in the literature[43].

***Computational details***

All the calculations are done in the MLatom ecosystem. In the case of learning the ethanol PES, all QM calculations were done via Gaussian 16[54] interfaced to MLatom[27, 28]. The ANI models were trained using TorchANI[55] interfaced to MLatom with the default settings (except that the models were trained on 90% of the labeled data set and validated on the rest of it). Geometry optimization and frequency calculations are done in MLatom through the interfaces to Gaussian and Atomic Simulation Environment (ASE)[56]. Further details of AL calculations for all applications are summarized in Table S1 (Supplementary Information).

**Code availability**

The code is available in the open-source MLatom (https://github.com/dralgroup/mlatom). The tutorials on how to perform the active learning simulations with MLatom is provided at https://xacs.xmu.edu.cn/docs/mlatom/tutorial_al.html.

**Author contributions**

P.O.D. conceived and designed the project and implemented an early version of active learning. Y.H. implemented the active learning procedure used in this manuscript, performed the tests, explored the procedure performance for the vibrational spectra simulations and conformer search, as well as required implementations of the initial conditions, MD, and assisted in reaction mechanism simulations. L.Z. performed the first implementations and calculations with active learning based on the related algorithms for a different problem and made contributions to the final active learning scripts. Q.Z. suggested and performed the exploration of the Diels–Alder reaction. F.G. made code improvements, e.g., optimized the performance in molecular dynamics. P.O.D. and Y.H. wrote the original version of the manuscript with input from all authors. All authors revised the manuscript. Y.H. and Q.Z. prepared all the figures.

**Acknowledgments**

P.O.D. acknowledges funding by the National Natural Science Foundation of China (No. 22003051 and funding via the Outstanding Youth Scholars (Overseas, 2021) project), the Fundamental Research Funds for the Central Universities (No. 20720210092), and via the Lab project of the State Key Laboratory of Physical Chemistry of Solid Surfaces. This project is supported by Science and Technology Projects of Innovation Laboratory for Sciences and Technologies of Energy Materials of Fujian Province (IKKEM) (No: RD2022070103).

**References**

1. Unke, O. T.; Chmiela, S.; Sauceda, H. E.; Gastegger, M.; Poltavsky, I.; Schutt, K. T.; Tkatchenko, A.; Muller, K. R., Machine Learning Force Fields. *Chem. Rev.* **2021,** *121*, 10142–10186.

2. Meuwly, M., Machine Learning for Chemical Reactions. *Chem. Rev.* **2021,** *121*, 10218–10239.

3. Zhang, L.; Ullah, A.; Pinheiro Jr, M.; Dral, P. O.; Barbatti, M., Chapter 15 - Excited-state dynamics with machine learning. In *Quantum Chemistry in the Age of Machine Learning*, Dral, P. O., Ed. Elsevier: 2023; pp 329–353.

4. Pios, S. V.; Gelin, M. F.; Ullah, A.; Dral, P. O.; Chen, L., Artificial-Intelligence-Enhanced On-the-Fly Simulation of Nonlinear Time-Resolved Spectra. *J. Phys. Chem. Lett.* **2024,** *15*, 2325–2331.

5. Chmiela, S.; Vassilev-Galindo, V.; Unke, O. T.; Kabylda, A.; Sauceda, H. E.; Tkatchenko, A.; Müller, K.-R., Accurate global machine learning force fields for molecules with hundreds of atoms. *Sci. Adv.* **2023,** *9*, eadf0873.

6. Gastegger, M.; Behler, J.; Marquetand, P., Machine learning molecular dynamics for the simulation of infrared spectra. *Chem. Sci.* **2017,** *8*, 6924–6935.

7. Smith, J. S.; Nebgen, B.; Lubbers, N.; Isayev, O.; Roitberg, A. E., Less is more: Sampling chemical space with active learning. *J. Chem. Phys.* **2018,** *148*, 241733.

8. Loeffler, T. D.; Patra, T. K.; Chan, H.; Cherukara, M.; Sankaranarayanan, S. K. R. S., Active Learning the Potential Energy Landscape for Water Clusters from Sparse Training Data. *J. Phys. Chem. C* **2020,** *124*, 4907–4916.

9. Smith, J. S.; Zubatyuk, R.; Nebgen, B.; Lubbers, N.; Barros, K.; Roitberg, A. E.; Isayev, O.; Tretiak, S., The ANI-1ccx and ANI-1x data sets, coupled-cluster and density functional theory properties for molecules. *Sci. Data* **2020,** *7*, 134.

10. Li, Z.; Kermode, J. R.; De Vita, A., Molecular dynamics with on-the-fly machine learning of quantum-mechanical forces. *Phys. Rev. Lett.* **2015,** *114*, 096405.

11. Jinnouchi, R.; Karsai, F.; Kresse, G., On-the-fly machine learning force field generation: Application to melting points. *Phys. Rev. B* **2019,** *100*, 014105.

12. Hu, D.; Xie, Y.; Li, X.; Li, L.; Lan, Z., Inclusion of Machine Learning Kernel Ridge Regression Potential Energy Surfaces in On-the-Fly Nonadiabatic Molecular Dynamics Simulation. *J. Phys. Chem. Lett.* **2018,** *9*, 2725–2732.

13. Botu, V.; Ramprasad, R., Adaptive machine learning framework to accelerate ab initio molecular dynamics. *Int. J. Quantum Chem.* **2015,** *115*, 1074–1083.

14. Vandermause, J.; Torrisi, S. B.; Batzner, S.; Xie, Y.; Sun, L.; Kolpak, A. M.; Kozinsky, B., On-the-fly active learning of interpretable Bayesian force fields for atomistic rare events. *npj Comput. Mater.* **2020,** *6*, 20.

15. Xie, Y.; Vandermause, J.; Ramakers, S.; Protik, N. H.; Johansson, A.; Kozinsky, B., Uncertainty-aware molecular dynamics from Bayesian active learning for phase transformations and thermal transport in SiC. *npj Comput. Mater.* **2023,** *9*, 36.

16. Wang, J.; Arantes, P. R.; Bhattarai, A.; Hsu, R. V.; Pawnikar, S.; Huang, Y. M.; Palermo, G.; Miao, Y., Gaussian accelerated molecular dynamics (GaMD): principles and applications. *WIREs Comput. Mol. Sci.* **2021,** *11*.

17. Laio, A.; Parrinello, M., Escaping free-energy minima. *Proc. Natl. Acad. Sci.* **2002,** *99*, 12562–12566.

18. Valsson, O.; Tiwary, P.; Parrinello, M., Enhancing Important Fluctuations: Rare Events and Metadynamics from a Conceptual Viewpoint. *Annu. Rev. Phys. Chem.* **2016,** *67*, 159–184.

19. Brezina, K.; Beck, H.; Marsalek, O., Reducing the Cost of Neural Network Potential Generation for Reactive Molecular Systems. *J. Chem. Theory Comput.* **2023,** *19*, 6589–6604.

20. Herr, J. E.; Yao, K.; McIntyre, R.; Toth, D. W.; Parkhill, J., Metadynamics for training neural network model chemistries: A competitive assessment. *J. Chem. Phys.* **2018,** *148*, 241710.

21. Bonati, L.; Parrinello, M., Silicon Liquid Structure and Crystal Nucleation from Ab Initio Deep Metadynamics. *Phys. Rev. Lett.* **2018,** *121*, 265701.

22. Kulichenko, M.; Barros, K.; Lubbers, N.; Li, Y. W.; Messerly, R.; Tretiak, S.; Smith, J. S.; Nebgen, B., Uncertainty-driven dynamics for active learning of interatomic potentials. *Nat. Comput. Sci.* **2023,** *3*, 230–239.

23. Seung, H. S.; Opper, M.; Sompolinsky, H., Query by committee. In *Proceedings of the fifth annual workshop on Computational learning theory*, Association for Computing Machinery: Pittsburgh, Pennsylvania, USA, 1992; pp 287–294.

24. Zhang, L.; Lin, D.-Y.; Wang, H.; Car, R.; E, W., Active learning of uniformly accurate interatomic potentials for materials simulation. *Phys. Rev. Mater.* **2019,** *3*.

25. Jospin, L. V.; Laga, H.; Boussaid, F.; Buntine, W.; Bennamoun, M., Hands-On Bayesian Neural Networks—A Tutorial for Deep Learning Users. *IEEE Comput.* **2022,** *17*, 29–48.

26. Esposito, C.; Landrum, G. A.; Schneider, N.; Stiefl, N.; Riniker, S., GHOST: Adjusting the Decision Threshold to Handle Imbalanced Data in Machine Learning. *J. Chem. Inf. Model.* **2021,** *61*, 2623–2640.

27. Dral, P. O.; Ge, F.; Hou, Y.-F.; Zheng, P.; Chen, Y.; Barbatti, M.; Isayev, O.; Wang, C.; Xue, B.-X.; Pinheiro Jr, M.; Su, Y.; Dai, Y.; Chen, Y.; Zhang, L.; Zhang, S.; Ullah, A.; Zhang, Q.; Ou, Y., MLatom 3: A Platform for Machine Learning-Enhanced Computational Chemistry Simulations and Workflows. *J. Chem. Theory Comput.* **2024,** *20*, 1193–1213.

28. Dral, P. O.; Ge, F.; Hou, Y.-F.; Zheng, P.; Chen, Y.; Xue, B.-X.; Pinheiro Jr, M.; Su, Y.; Dai, Y.; Chen, Y.; Zhang, S.; Zhang, L.; Ullah, A.; Ou, Y. *MLatom: A Package for Atomistic Simulations with Machine Learning*, Xiamen University, Xiamen, China, http://MLatom.com, 2013–2024.

29. Hou, Y.-F.; Ge, F.; Dral, P. O., Explicit Learning of Derivatives with the KREG and pKREG Models on the Example of Accurate Representation of Molecular Potential Energy Surfaces. *J. Chem. Theory Comput.* **2023,** *19*, 2369–2379.

30. Pinheiro, M.; Ge, F.; Ferré, N.; Dral, P. O.; Barbatti, M., Choosing the right molecular machine learning potential. *Chem. Sci.* **2021,** *12*, 14396–14413.

31. Zhang, S.; Makoś, M. Z.; Jadrich, R. B.; Kraka, E.; Barros, K.; Nebgen, B. T.; Tretiak, S.; Isayev, O.; Lubbers, N.; Messerly, R. A.; Smith, J. S., Exploring the frontiers of condensed-phase chemistry with a general reactive machine learning potential. *Nat. Chem.* **2024**.

32. Devereux, C.; Smith, J. S.; Huddleston, K. K.; Barros, K.; Zubatyuk, R.; Isayev, O.; Roitberg, A. E., Extending the Applicability of the ANI Deep Learning Molecular Potential to Sulfur and Halogens. *J. Chem. Theory Comput.* **2020,** *16*, 4192–4202.

33. Cortes, C.; Jackel, L. D.; Solla, S. A.; Vapnik, V.; Denker, J. S., Learning curves: Asymptotic values and rate of convergence. In *Advances in Neural Information Processing Systems*, Morgan Kaufmann Publishers: San Mateo, CA, 1994; pp 327–334.

34. Leys, C.; Ley, C.; Klein, O.; Bernard, P.; Licata, L., Detecting outliers: Do not use standard deviation around the mean, use absolute deviation around the median. *J. Exp. Soc. Psychol.* **2013,** *49*, 764–766.

35. Zhang, L.; Hou, Y.-F.; Ge, F.; Dral, P. O., Energy-conserving molecular dynamics is not energy conserving. *Phys. Chem. Chem. Phys.* **2023,** *25*, 23467–23476.

36. Tokita, A. M.; Behler, J., How to train a neural network potential. *J. Chem. Phys.* **2023,** *159*, 121501.

37. Miksch, A. M.; Morawietz, T.; Kästner, J.; Urban, A.; Artrith, N., Strategies for the construction of machine-learning potentials for accurate and efficient atomic-scale simulations. *Mach. Learn. Sci. Technol.* **2021,** *2*, 031001.

38. Morrow, J. D.; Gardner, J. L. A.; Deringer, V. L., How to validate machine-learned interatomic potentials. *J. Chem. Phys.* **2023,** *158*, 121501.

39. Conte, R.; Houston, P. L.; Qu, C.; Li, J.; Bowman, J. M., Full-dimensional, ab initio potential energy surface for glycine with characterization of stationary points and zero-point energy calculations by means of diffusion Monte Carlo and semiclassical dynamics. *J. Chem. Phys.* **2020,** *153*, 244301.

40. Csaszar, A. G., Conformers of gaseous glycine. *J. Am. Chem. Soc.* **1992,** *114*, 9568–9575.

41. Sewell, T. D.; Thompson, D. L., Classical Trajectory Methods for Polyatomic Molecules. *Int. J. Mod. Phys. B* **1997,** *11*, 1067–1112.

42. Zhang, Y.; Cao, C.; She, Y.; Yang, Y. F.; Houk, K. N., Molecular Dynamics of Iron Porphyrin-Catalyzed C-H Hydroxylation of Ethylbenzene. *J. Am. Chem. Soc.* **2023,** *145*, 14446–14455.

43. Black, K.; Liu, P.; Xu, L.; Doubleday, C.; Houk, K. N., Dynamics, transition states, and timing of bond formation in Diels-Alder reactions. *Proc. Natl. Acad. Sci.* **2012,** *109*, 12860–12865.

44. Tyukina, S. P.; Velmiskina, J. A.; Dmitrienko, A. O.; Medvedev, M. G., Binomial Uncertainty in Molecular Dynamics-Based Reactions Analysis. *J. Phys. Chem. Lett.* **2024**, 2105–2110.

45. Hou, Y.-F.; Zhang, Q.; Dral, P. O., Surprising new dynamics phenomena in Diels–Alder reaction of $C_{60}$ uncovered with AI. *ChemRxiv* **2024**.

46. Barbatti, M.; Bondanza, M.; Crespo-Otero, R.; Demoulin, B.; Dral, P. O.; Granucci, G.; Kossoski, F.; Lischka, H.; Mennucci, B.; Mukherjee, S.; Pederzoli, M.; Persico, M.; Pinheiro Jr, M.; Pittner, J.; Plasser, F.; Sangiogo Gil, E.; Stojanovic, L., Newton-X Platform: New Software Developments for Surface Hopping and Nuclear Ensembles. *J. Chem. Theory Comput.* **2022,** *18*, 6851–6865.

47. Doubleday, C.; Bolton, K.; Hase, W. L., Direct Dynamics Quasiclassical Trajectory Study of the Thermal Stereomutations of Cyclopropane. *J. Phys. Chem. A* **1998,** *102*, 3648–3658.

48. McQuarrie, D. A., *Statistical Thermodynamics*. University Science Books: CA: 1973.

49. Chapman, S.; Bunker, D. L., An exploratory study of reactant vibrational effects in CH3 + H2 and its isotopic variants. *J. Chem. Phys.* **1975,** *62*, 2890–2899.

50. Sloane, C. S.; Hase, W. L., On the dynamics of state selected unimolecular reactions: Chloroacetylene dissociation and predissociation. *J. Chem. Phys.* **1977,** *66*, 1523–1533.

51. Cho, Y. J.; Vande Linde, S. R.; Zhu, L.; Hase, W. L., Trajectory studies of SN2 nucleophilic substitution. II. Nonstatistical central barrier recrossing in the Cl−+CH3Cl system. *J. Chem. Phys.* **1992,** *96*, 8275–8287.

52. Bunker, D. L., Classical Trajectory Methods. In *Methods in Computational Physics: Advances in Research and Applications*, Alder, B.; Fernbach, S.; Rotenberg, M., Eds. Elsevier: 1971; Vol. 10, pp 287–325.

53. Pracht, P.; Grant, D. F.; Grimme, S., Comprehensive Assessment of GFN Tight-Binding and Composite Density Functional Theory Methods for Calculating Gas-Phase Infrared Spectra. *J. Chem. Theory Comput.* **2020,** *16*, 7044−7060.

54. Frisch, M. J.; Trucks, G. W.; Schlegel, H. B.; Scuseria, G. E.; Robb, M. A.; Cheeseman, J. R.; Scalmani, G.; Barone, V.; Petersson, G. A.; Nakatsuji, H.; Li, X.; Caricato, M.; Marenich, A. V.; Bloino, J.; Janesko, B. G.; Gomperts, R.; Mennucci, B.; Hratchian, H. P.; Ortiz, J. V.; Izmaylov, A. F.; Sonnenberg, J. L.; Williams; Ding, F.; Lipparini, F.; Egidi, F.; Goings, J.; Peng, B.; Petrone, A.; Henderson, T.; Ranasinghe, D.; Zakrzewski, V. G.; Gao, J.; Rega, N.; Zheng, G.; Liang, W.; Hada, M.; Ehara, M.; Toyota, K.; Fukuda, R.; Hasegawa, J.; Ishida, M.; Nakajima, T.; Honda, Y.; Kitao, O.; Nakai, H.; Vreven, T.; Throssell, K.; Montgomery Jr., J. A.; Peralta, J. E.; Ogliaro, F.; Bearpark, M. J.; Heyd, J. J.; Brothers, E. N.; Kudin, K. N.; Staroverov, V. N.; Keith, T. A.; Kobayashi, R.; Normand, J.; Raghavachari, K.; Rendell, A. P.; Burant, J. C.; Iyengar, S. S.; Tomasi, J.; Cossi, M.; Millam, J. M.; Klene, M.; Adamo, C.; Cammi, R.; Ochterski, J. W.; Martin, R. L.; Morokuma, K.; Farkas, O.; Foresman, J. B.; Fox, D. J. *Gaussian 16 Rev. B.01*, Wallingford, CT, 2016.

55. Gao, X.; Ramezanghorbani, F.; Isayev, O.; Smith, J. S.; Roitberg, A. E., TorchANI: A Free and Open Source PyTorch-Based Deep Learning Implementation of the ANI Neural Network Potentials. *J. Chem. Inf. Model.* **2020,** *60*, 3408–3415.

56. Hjorth Larsen, A.; Jørgen Mortensen, J.; Blomqvist, J.; Castelli, I. E.; Christensen, R.; Dułak, M.; Friis, J.; Groves, M. N.; Hammer, B.; Hargus, C.; Hermes, E. D.; Jennings, P. C.; Bjerre Jensen, P.; Kermode, J.; Kitchin, J. R.; Leonhard Kolsbjerg, E.; Kubal, J.; Kaasbjerg, K.; Lysgaard, S.; Bergmann Maronsson, J.; Maxson, T.; Olsen, T.; Pastewka, L.; Peterson, A.; Rostgaard, C.; Schiøtz, J.; Schütt, O.; Strange, M.; Thygesen, K. S.; Vegge, T.; Vilhelmsen, L.; Walter, M.; Zeng, Z.; Jacobsen, K. W., The atomic simulation environment-a Python library for working with atoms. *J. Phys.: Condens. Matter* **2017,** *29*, 273002.